  \documentclass[conference]{IEEEtran}

\usepackage[colorlinks,urlcolor=blue,linkcolor=blue,citecolor=blue]{hyperref}

\usepackage{amsmath}
\usepackage{amssymb}
\usepackage{amsfonts}
\usepackage{graphicx}
\usepackage{url}
\usepackage{cite}
\usepackage{mathtools}
\usepackage{cuted}
\usepackage{etoolbox}
\usepackage{subcaption}
\usepackage{tikz}
 \usepackage{diagbox}
\usepackage{multirow}
\usepackage{mathrsfs}
\usepackage{booktabs} 
\usepackage{subcaption}
\usepackage{float}

\IEEEoverridecommandlockouts
\begin{document}

\title{RSCNet: Dynamic CSI Compression for Cloud-based WiFi Sensing}
\author{
\IEEEauthorblockN{Borna Barahimi\IEEEauthorrefmark{1}, Hakam Singh\IEEEauthorrefmark{1}, Hina Tabassum\IEEEauthorrefmark{1}, Omer Waqar\IEEEauthorrefmark{2}, Mohammad Omer\IEEEauthorrefmark{3} \thanks{This research was supported by NSERC Alliance Grants funded by Cognitive Systems Inc.  and the Natural Sciences and Engineering Research Council (NSERC) of Canada.}}
\IEEEauthorblockA{\IEEEauthorrefmark{1}Electrical Engineering and Computer Science Department, York University, Canada. \\\IEEEauthorrefmark{2} School of Computing, University of the Fraser Valley, BC, Canada. \\\IEEEauthorrefmark{3} Cognitive Systems Corp., Waterloo, Canada. }
}

\raggedbottom

\maketitle	
\begin{abstract}
 WiFi-enabled Internet-of-Things (IoT) devices are evolving from mere communication devices to sensing instruments, leveraging Channel State Information (CSI) extraction capabilities. Nevertheless, resource-constrained IoT devices and the intricacies of deep neural networks necessitate transmitting CSI  to cloud servers for sensing. Although feasible, this leads to considerable communication overhead. In this context, this paper develops a novel  Real-time Sensing and Compression Network (RSCNet) which enables sensing with compressed CSI; thereby reducing the communication overheads. RSCNet facilitates optimization across CSI windows composed of a few CSI frames. Once transmitted to cloud servers, it employs Long Short-Term Memory (LSTM) units to harness data from prior windows, thus bolstering both the sensing accuracy and CSI reconstruction. RSCNet adeptly balances the trade-off between CSI compression and sensing precision, thus streamlining real-time cloud-based WiFi sensing with reduced communication costs. Numerical findings demonstrate the gains of RSCNet over the existing benchmarks like SenseFi, showcasing a sensing accuracy of 97.4\% with minimal CSI reconstruction error. Numerical results also show a computational analysis of the proposed RSCNet as a function of the number of CSI frames.

\end{abstract}


\section{Introduction}
With the rapid increase in 6G-enabled \cite{6g} or WiFi-enabled Internet of Things (IoT) devices, wireless sensing can redefine the sensing paradigm. WiFi sensing places emphasis on privacy and facilitates ubiquitous, non-invasive sensing as users do not need to carry  sensors \cite{9831898}. WiFi sensing is cost-efficient as it capitalizes on existing WiFi infrastructure. Unlike systems limited by line-of-sight (LOS) constraints, wireless signals provide rich data through reflection and diffraction, even in non-line-of-sight (NLOS) scenarios where obstructions exist between the target and WiFi devices. Notably, these signals can operate in dark, offering a round-the-clock functionality that cameras cannot match without infrared (IR) illuminators. WiFi signals permit the extraction of specific details, such as human position and vital signs, without visual information.  WiFi sensing enables localization \cite{8397121}, human activity recognition (HAR) \cite{9834923}, gesture recognition \cite{9233449}, and  respiration detection \cite{9831898}. 
WiFi sensing relies heavily on extracting accurate Channel State Information (CSI) \cite{10.1145/1925861.1925870,yang2023sensefi}.
CSI captures variations in radio frequency (RF) signals as they move through a physical space, interacting with objects or human bodies, causing phenomena like reflection, diffraction, and scattering. These interactions result in multipath effects, which convey valuable information about the surrounding environment \cite{9900419}. 

Recently, a variety of research works have demonstrated the significance of deep learning methods for WiFi sensing. Zhuravchak et al. \cite{zhuravchak2022human} proposed a Long Short-Term Memory (LSTM)-based method for HAR, WiGRUNT\cite{9759238} leverages ResNet architecture in conjunction with a dual attention mechanism for gesture recognition. Moreover, AFEE-MatNet \cite{9834923} and Widar3.0\cite{9516988} addressed the environment dependency issue of deep learning methods through extensive preprocessing and complex neural networks.

Nevertheless, all aforementioned methods  demand substantial computation resources. In practice, IoT devices at the edge  suffer from limited power and computation capabilities. To overcome this issue, transmission of CSI to cloud servers becomes necessary to enable cloud-based WiFi sensing. Yet, high dimensionality and sampling rate of CSI lead to a substantial transmission overhead \cite{9797871}.   
As such, there is an undeniable need to compress the CSI at the edge  before its transmission to the cloud. In addition, besides sensing, CSI reconstruction is becoming essential for data logging at cloud servers in systems like Healthcare IoT \cite{7207365}. Thus, cloud servers need not only host  accurate sensing but also CSI reconstruction~\cite{yang2022efficientfi}.

None of the aforementioned research works considered the problem of joint CSI compression and sensing.  Recently, EfficientFi \cite{yang2022efficientfi} tackled  sensing and CSI compression jointly by compressing CSI into a quantized low-dimensional space at the edge using a trainable shared codebook between the access point (AP) and the cloud server. However, EfficientFi is vulnerable to high communication overhead as their decoder in the cloud requires the indices from the max-pool layers of the encoder to up-sample the CSI to its original dimension.

Unlike existing works, we develop a novel architecture, \textbf{R}eal-time \textbf{S}ensing and \textbf{C}ompression \textbf{Net}work (RSCNet), that leverages CSI windowing and recurrent blocks to address the problem of joint CSI compression and sensing.
Our contributions are summarized as follows:
\begin{enumerate}
    \item We present RSCNet, a novel network architecture designed  specifically for real-time cloud-based Wi-Fi sensing for HAR. The network encompasses an encoder for CSI compression at the edge and a dual-network in the cloud for both WiFi sensing and CSI reconstruction.
The proposed architecture offers a lightweight encoder by using dilated convolutional layers and residual connections to be employable in low-resource edge devices and an expandable decoder design to optimize the trade-off between complexity and reconstruction error. 
    \item We demonstrate the significance of window-based CSI compression, resulting in efficient real-time HAR with reduced communication overheads compared to traditional fixed task duration based samplings.
     \item   We incorporate LSTM-based recurrent blocks to leverage the time-series representation of CSI windows, improving compression and human activity classification.
     \item  Through extensive experimentation on UT-HAR dataset \cite{8067693}, we observe that a window size of 50 frames strikes a balance, yielding a peak HAR accuracy of 97.2\% and  NMSE of -22.759~dB, under a compression ratio of \(\eta = 1/90\). The framework maintains its robust performance across a range of compression ratios. We note that expanding the decoder with an expansion rate of 5 results in a reduction of reconstruction error by 0.84 dB compared to an expansion rate of 1, albeit at the cost of approximately 9 times higher Floating Point Operations Per Second (FLOP) counts.
     \item We have made RSCNet's source code available on GitHub at \url{https://github.com/bornabr/RSCNet}.  
\end{enumerate}

\section{Fundamentals of CSI in WiFi Sensing}

CSI  characterizes the way wireless signals propagate from a transmitter to a receiver over specific carrier frequencies. It offers a detailed description of the wireless environment, highlighting how it is influenced by multi-path effects. These effects emerge from variations and movements of the transmitter, receiver, and nearby objects.

The CSI, represented as \(H\), defines the transformation of the transmitted signal \(x\) to produce the received signal \(y\), accounting for noise \(\eta\) at a given timestamp \(t\), i.e.,
$
y = Hx + \eta.
$
CSI can be represented using the Channel Impulse Response (CIR), i.e., $h (\tau) = \sum^N_{n=1} a_n \delta(t - \tau_n)$, where the factors \(a_n \) and \(\tau_n(t)\) describe the amplitude attenuation and propagation delay of the \(n\)th path at timestamp \(t\), respectively. $\delta(t)$ denotes the Dirac-delta function. Subsequently,  the channel frequency response (CFR) at carrier frequency \(f\) and time \(t\) can be modelled as follows:
\begin{equation}
H(f;t) = \sum^N_{n=1} a_n  e^{-j2 \pi f \tau_n (t)},
\end{equation}
To measure CSI, the WiFi transmitter emits known Long Training Symbols (LTFs). Using these symbols and the received signals, the receiver calculates the CSI matrix \cite{10.1145/3310194}.
The CSI  for each subcarrier \(i\) at timestamp \(t\) can be modeled as a complex number
$
H_{f_i,t} = |H_{f_i,t}|e^{j\angle H_{f_i,t}}.
$
Both the amplitude \(|H_{f_i,t}|\) and phase \(\angle H_{f_i,t}\) are impacted by the displacements and
movements of the transmitter, receiver, and surrounding objects and humans. In contrast to the received signal strength, CSI offers superior resolution for sensing tasks. This is attributed to its precise characterization of phase shift and amplitude attenuation for each subcarrier. This comprehensive information allows CSI to serve as \textit{WiFi image frames} of the environment, providing a nuanced understanding \cite{yang2023sensefi}.

In WiFi sensing, a sequence of \textit{CSI frames} compiles into a matrix \(H = \{ H_t \in \mathbb{C}^{N_a \times N_s} \}\). Here, \(N_a\) and \(N_s\) represent the number of antennas and subcarriers, respectively, while \(N_t\), the length of \(H\), stands for the time dimension. 
By segmenting \(H\) over time dimension, it becomes possible to generate a series of distinct windows, with each window encompassing a given number of CSI frames \(N_f\). In this paper, these specific constructs will be consistently referred to as \textit{CSI window}. For WiFi sensing applications, typically the amplitude information \(|H_t|\) is deemed sufficient. Thus, RSCNet prioritizes encoding and decoding the amplitude \(|H|\) for sensing applications.

\section{RSCNet: Joint CSI Compression and Reconstruction}
\label{sec:methodology}

\begin{figure*}[ht]
  \centering
  \includegraphics[width=\linewidth]{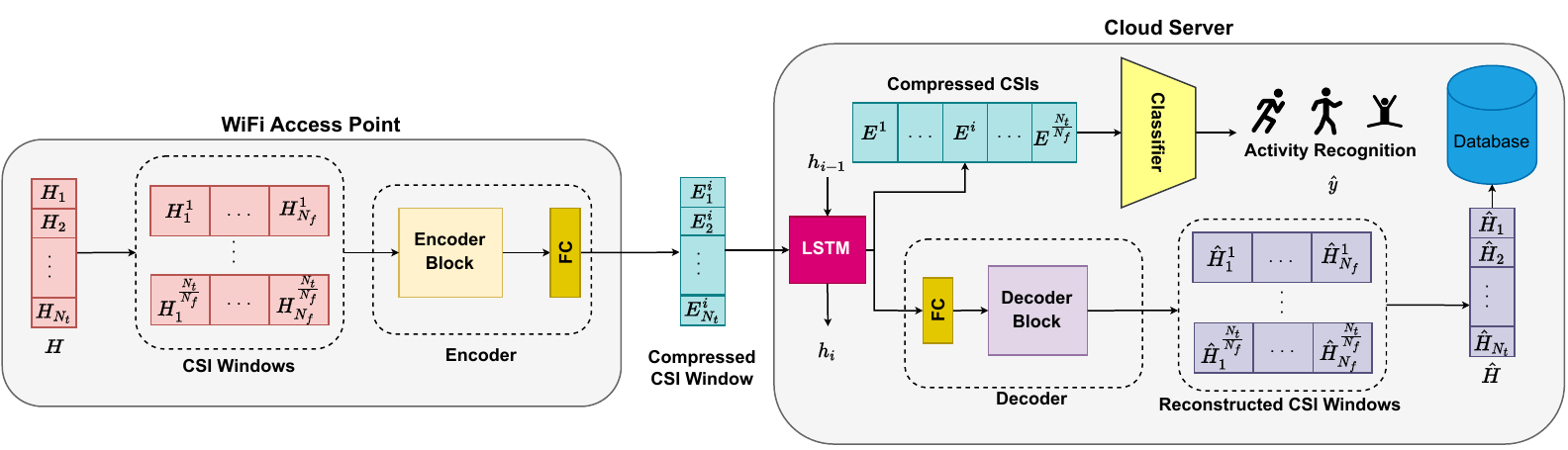}
  \caption{Design of the proposed RSCNet system}
  \label{fig:network}
\end{figure*}

In this section, we discuss our proposed framework, RSCNet, and give details on the real-time compression of CSI windows and the multi-task learning process of CSI reconstruction and HAR  as demonstrated in Fig.~\ref{fig:network}.

\subsection{System Overview}

Our model, RSCNet, is bifurcated into the edge and cloud models. The edge model is employed at the edge devices or access points (APs) where CSI is acquired. A CSI window, which contains several frames of CSI undergoes compression through an encoder. Compressed CSI embedding is then transmitted to the server where the cloud model, which incorporates recurrent blocks for temporal CSI enhancement, is used to do CSI reconstruction using a decoder. Additionally, a classifier accumulates the outputs of recurrent blocks for all of the compressed CSI windows of an activity to do sensing.

The encoder and decoder are inspired by DCRNet \cite{9797871}. The encoder has an encoder block for feature extraction and a fully connected layer (FC) which receives the flattened version of the output of the encoder block to reduce the data size based on the compression rate ($\eta$). Similarly, the decoder has an FC layer to bring the size of data back to its original one and reshape it to make the CSI matrix then it goes to a decoder block which aims to restore the original CSI. This architecture utilizes dilated convolutional layers to enable large respective fields while keeping a minimal number of parameters, ensuring a lightweight model design suitable for edge devices. Moreover, this decoder design provides a trade-off between the decoder's number of parameters i.e. computational cost and its reconstruction capabilities via an expansion rate hyperparameter, $\rho$.

Other than CSI reconstruction, the cloud model performs HAR  using a multi-layer perceptron (MLP) as the classifier. RSCNet's objective is to achieve high HAR accuracy  with minimal error in CSI reconstruction due to compression. Consequently, RSCNet is capable of compressing CSI to a low-dimension space, without diminishing the HAR accuracy.

\subsection{CSI Windowing for Real-Time Compression and HAR}

RSCNet utilizes segmentation of the continuous CSI data stream, represented as $N_a \times N_s \times N_t$, along the time dimension into distinct windows of size $N_a \times N_s \times N_f$, where $N_f \leq N_t$ is the number of CSI frames in each window. This segmentation enables the system to manage data more effectively, facilitating both real-time HAR and efficient compression. After feature extraction in the edge model, the output is compressed, with the compression ratio, $\eta \in [\frac{1}{N_a \cdot N_s \cdot N_f}, 1] $, determining the degree to which the CSI window size is reduced. For instance, $\eta = 1/90$ means that the compressed CSI window is 90 times smaller than the original CSI window.

In the cloud, each compressed CSI window undergoes the recurrent block to extract features related to the previous CSI windows. Then, it goes through the reconstruction process. Once a CSI window is restored, it gets merged with other windows to regenerate the complete original CSI activity sample. For classification/HAR, all compressed CSI windows are stacked to form an embedding that captures the essence of the original CSI, denoted as $(N_a \cdot N_s \cdot N_h \cdot \eta) \times (N_t/N_f)$, where $N_h$ is the hidden state size of the recurrent block. This data is subsequently flattened and given to the classifier for HAR.
The RSCNet framework, by strategically handling CSI data in real-time, ensures computational efficiency, minimizes latency and preserves data's temporal fidelity. 

\subsection{Recurrent Block Integration}

To further optimize the efficiency and accuracy of CSI reconstruction and HAR, we incorporate a recurrent block at the beginning of the cloud model. This block utilizes LSTM units, well-recognized for their powers in capturing temporal relationships in sequential data. Each LSTM receives its input from two sources: the current compressed CSI window and the hidden state from the preceding CSI window. This allows the LSTM to benefit from the temporal continuity inherent in sequential CSI windows, enhancing its understanding of the current window in the context of previous data.

After processing through the LSTM, its output serves a dual purpose. Firstly, it's fed into the decoder to refine the CSI reconstruction, ensuring it is both accurate and contextually relevant. Secondly, when combined with the outputs from other windows, it forms a comprehensive representation fed into the classifier. This ensemble of temporal information improves the classifier's accuracy, allowing for more precise WiFi-enable HAR or sensing.

\subsection{Encoder and Decoder Designs}

The encoder block has a $5\times5$ head convolution at the beginning that extracts features from the input CSI matrix and fuses the information from different antennas. The data goes through a residual network with three asymmetric dilated convolution layers, each with a $3\times3$ kernel size, namely DConv blocks. In contrast to standard convolutions, the DConv layers implement dilated convolutions to enhance the receptive field of the convolutional layers without necessitating an increase in the kernel size, ensuring computational efficiency is maintained. Specifically, DConv extracts features at a particular interval, denoted as $d$ or the dilation rate. Mathematically, the two-dimensional dilated convolution operation, excluding bias, can be expressed as:

\begin{align} (\boldsymbol{I} * \boldsymbol{K})[i,j]=\sum _{m}\sum _{n} {\boldsymbol{I}[i+dm,j+dn]\cdot \boldsymbol{K}[m,n]}, \end{align}

where $*$ denotes the dilated convolution operation, $\mathbf{I}$ represents the two-dimensional input, and $\mathbf{K}$ is the convolution kernel. Indices $m$ and $n$ traverse the spatial dimensions of the kernel $\mathbf{K}$. The dilation operation effectively increases the size of the kernel to $
k' = k + (k - 1)(d - 1),
$ with $k$ being the original kernel size and $k'$ representing the effective kernel size after dilation which is bigger than $k$ demonstrating the larger receptive field of DConv. Each DConv layer in our model utilizes a unique dilation rate, $d$, allowing for varying receptive fields and a richer feature extraction from the input CSI matrix. These features are concatenated with a separate standard $3\times3$ convolution. After concatenation, a $1\times1$ convolution is used to reshape the data back to its original shape form before adding the initial input as residuals to the final output.

The decoder block does initial feature extraction using a $5\times5$ head convolution. The decoder employs two sequential dilated residual decoder blocks to recover the compressed information. The residual decoder blocks follow the design principle of the encoder, using two parallel branches and an identity map. The first branch uses a $3\times3$ dilated convolution with $d=2$ to increase the feature dimension based on $\rho$, a flexible expansion rate hyperparameter, which can be adjusted according to devices with varying computational capacities. $3\times1$ and $1\times3$ convolution layers with dilation of 3 and out channel size of $c = 3\rho$, where $\rho$ is the expansion rate of the model, are used. A $3\times3$ convolution layer is used to reduce the channel dimension back to $N_a$. In the second branch, convolution layers filters of sizes $1\times3$ and $3\times1$ are used to increase and reduce feature dimensions at the beginning and end of the branch, respectively. $5\times1$ and $1\times5$ convolutions follow a similar out channel size as in the first branch, but without any dilation. Similar to the encoder, the output of these branches is concatenated and a $1\times1$ convolution is used to reshape the data back to its original dimensions.

For an in-depth understanding of the encoder and decoder blocks, readers can refer to the DCRNet architecture in \cite{9797871}.

\subsection{Multi-task Learning for HAR and CSI Reconstruction}

Within the architecture of RSCNet, the compressed CSI output from the encoder plays a dual role: it is used both for CSI reconstruction and enabling HAR. To ensure that this compressed representation is both a faithful representation of the original CSI and retains the necessary discriminative features for HAR, a unique loss function, $\mathcal{L}$, is employed:
\begin{equation}
\mathcal{L} =  \mathcal{L}_{c} + \lambda \mathcal{L}_{r}
\label{eq:loss}
\end{equation}
Here, $\mathcal{L}_{r}$ represents the Mean Square Error (MSE) which captures the error in CSI reconstruction. It's mathematically expressed in Equation \ref{eq:mse}:
\begin{equation}
\mathcal{L}_{r}=\mathbb {E}\left\lbrace {\left\Vert \boldsymbol{H}-\hat{\boldsymbol{H}}\right\Vert _{2}^{2}}\right\rbrace
\label{eq:mse}
\end{equation}
In the above, $\boldsymbol{H}$ represents the original CSI, while $\hat{\boldsymbol{H}}$ denotes its reconstructed counterpart. The MSE seeks to minimize the differences between these two matrices, aiming for accurate CSI reconstruction with minimal error.

On the other hand, $\mathcal{L}_{c}$ signifies the Cross-Entropy loss for HAR, as detailed in Equation \ref{eq:cross_entropy}:
\begin{equation}
\mathcal{L}_{c}(x, y) = -\mathbb{E}_{(x, y)} \sum_{i=1}^{C} \left [ \mathbb{I}[y = i] \log \left( \sigma \left( \hat{y}_{i}(x) \right) \right) \right]
\label{eq:cross_entropy}
\end{equation}
In this formulation, $\mathbb{I}[y = i]$ is an indicator function that returns $1$ if the label $y$ is equal to class $i$ and $0$ otherwise. $\sigma$ is the softmax function, and $\hat{y}_{i}(x)$ provides the predicted probability of class $i$ for a given input $x$ by the classifier. The Cross-Entropy loss calculates the difference between the predicted probabilities and actual class labels, with a focus on optimizing sensing accuracy.

The coefficient $\lambda$ in the combined loss function serves as a weight factor, balancing the scale of two loss functions. It ensures that while the compressed CSI is a robust representative of the original data, it also retains the nuances necessary for precise HAR or sensing.

\section{Experiment and Evaluation}

In this section, we  explain the data set-up followed by the settings for model training and  evaluation criteria. Then, we present NMSE, HAR accuracy, and FLOPs count of RSCNet and compare our results with benchmarks like SenseFi.
\subsection{Experimental Settings}

\subsubsection{\textbf{Data Setup}}
In order to validate the efficiency of the RSCNet framework, we utilized the UT-HAR dataset \cite{8067693}, collected by the University of Toronto. This dataset, gathered using Intel 5300 NIC \cite{10.1145/1925861.1925870}, comprises 3 pairs of antennas and 30 subcarriers per pair. Given that the UT-HAR dataset originally consisted of continuous CSI data, we opt for a segmented version offered by \cite{yang2023sensefi}, containing approximately 5000 samples, each with 250 frames of CSI matrices. This HAR dataset encompasses 7 distinct categories for human activity: lie down, fall, walk, run, sit down, stand up, and empty room. For the purpose of our experiments, we divide the dataset into training, validation, and testing sets, consisting of 3977, 496, and 500 samples, respectively.

\subsubsection{\textbf{Training Setting}}
We implement our model using the PyTorch framework. The model is optimized using the Stochastic Gradient Descent (SGD) with a learning rate of 0.01, momentum of 0.9, and weight decay of $1.5 \times 10^6$. For learning rate adjustment during training, we utilize the Cosine Annealing schedule for learning rate adjustments. The training process is conducted with a batch size of 512 across 300 epochs. Additionally, we note that the scales of $\mathcal{L}_{c}$ and $\mathcal{L}_{r}$ become similar and yield the best classification accuracy and reconstruction error when using $\lambda = 50$. The expansion rate, used for the decoder block is $\rho=1$ unless specified otherwise. Finally, the classifier utilized for HAR  has two hidden layers containing 512 and 128 nodes.

\subsubsection{\textbf{Evaluation Criterion}}
To assess the sensing capabilities of the framework, we evaluate the recognition accuracy on the test dataset. The compression performance is ascertained using the Normalized Mean Square Error (NMSE), which is reported in dB and defined as follows:

\begin{equation}
\text{NMSE}=\mathbb {E}\left\lbrace {\frac{\left\Vert \boldsymbol{H}-\hat{\boldsymbol{H}}\right\Vert _{2}^{2}}{\left\Vert \boldsymbol{H}\right\Vert _{2}^{2}}}\right\rbrace
\label{eq:nmse}
\end{equation}

To further comprehend our model's efficiency, particularly in resource-limited scenarios, we assessed its complexity by computing FLOP counts, underscoring the model's suitability for resource-constrained IoT devices and cloud servers.

\subsection{Performance of the RSCNet}

\subsubsection{Choice of Number of CSI Frames 
 $N_f$} RSCNet gives flexibility in choosing the number of CSI frames in a window, $N_f$, for compression and transmission to the cloud. The selection of $N_f$ is an important decision as it can affect the complexity of the encoder and the decoder, the frequency of CSI transmission, and the HAR performance. In the results depicted in Table \ref{table:frames}, we observe that by increasing the number of frames, $N_f$, from 5 to 50, there's an overall improvement in HAR accuracy, peaking at 97.2\%. However, when extending the $N_f$ further to a maximum of 250 which is the size of the time dimension, $N_t$, the HAR accuracy reduces and settles at 95\%. It becomes evident that neither a minimal nor an exceedingly large $N_f$ is optimal. While the former might lack capturing sufficient temporal features required for sensing recognition, the latter reduces the number of windows or sequence length which impacts the power of recurrent block to capture temporal information.

\subsubsection{FLOPs vs Number of CSI Frames $N_f$} 
\begin{figure}[htbp]
    \centering
    \includegraphics[width=\linewidth]{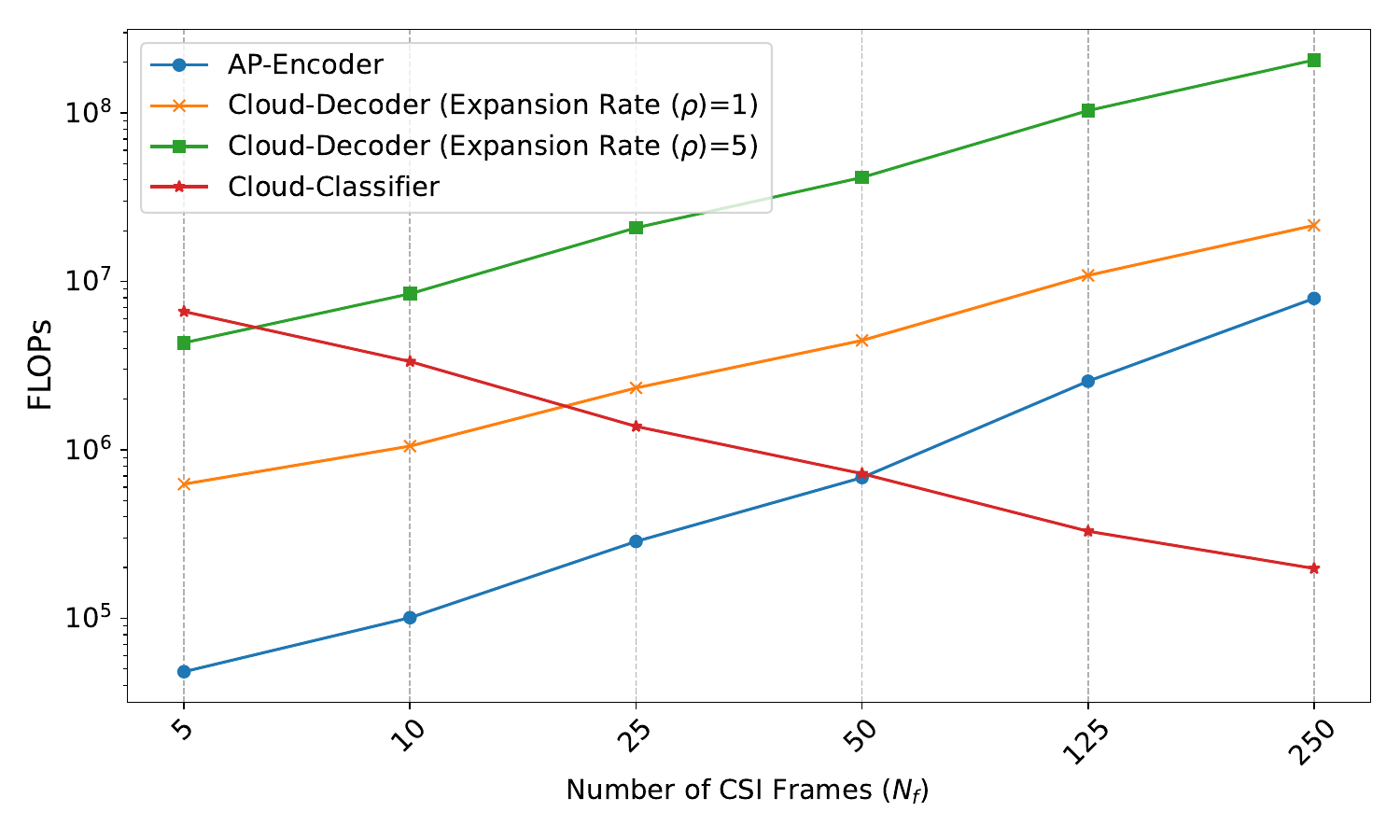}
    \caption{Comparative FLOP counts analysis for different CSI frame numbers with $\eta = 1/90$ compression}
    \label{fig:flops}
\end{figure}

The complexity of the network is also influenced by $N_f$, as illustrated in Fig.~\ref{fig:flops}. Given that the data shape remains consistent across most network layers, an increase in the number of frames can amplify the encoder's complexity by up to 150 times and the decoder's by up to 50 times. However, the classifier complexity follows a reverse pattern as its number of parameters depends on the sequence length which decreases as the number of CSI frames increases. Thus, opting for a smaller window size can enhance the encoder's efficiency, but it will increase the classifier complexity in the cloud.

Furthermore, as $N_f$ increases in a window, both the size of the CSI data and its compressed version expand, concurrently leading to a reduced frequency of data transmission but with higher overhead. This dynamic highlights the trade-off RSCNet faces between window size and transmission frequency. Therefore, an improper $N_f$ selection can entail transmission overheads, either as frequent transmissions for smaller $N_f$ or as data-heavy, infrequent bursts for larger ones. Due to the balance achieved at $N_f=$50 frames, in terms of HAR accuracy, NMSE, and complexity, we chose this frame count for subsequent tests. This selection underscores the need to optimize $N_f$ to ensure efficient data compression, optimal transmission, and sustained HAR accuracy. 

\begin{table}[htbp]
\centering
\caption{Comparing NMSE and accuracy across different frame counts with $\eta = 1/90$ compression}
\begin{tabular}{c|cc}
\toprule
$N_f$ & NMSE (dB) & Accuracy (\%) \\
\midrule
5   & \underline{-23.212} & 91.80  \\
10  & \textbf{-23.915} & 95.80 \\
25  & -22.373 & 95.60 \\
50  & -22.759 & \textbf{97.20} \\
125 & -21.881 & \underline{96.20} \\
250 & -20.160 & 95.00 \\
\bottomrule
\end{tabular}
\label{table:frames}
\end{table}

\begin{figure}[htb]
    \centering
    \begin{subfigure}{0.45\textwidth}
        \centering
        \includegraphics[width=\linewidth]{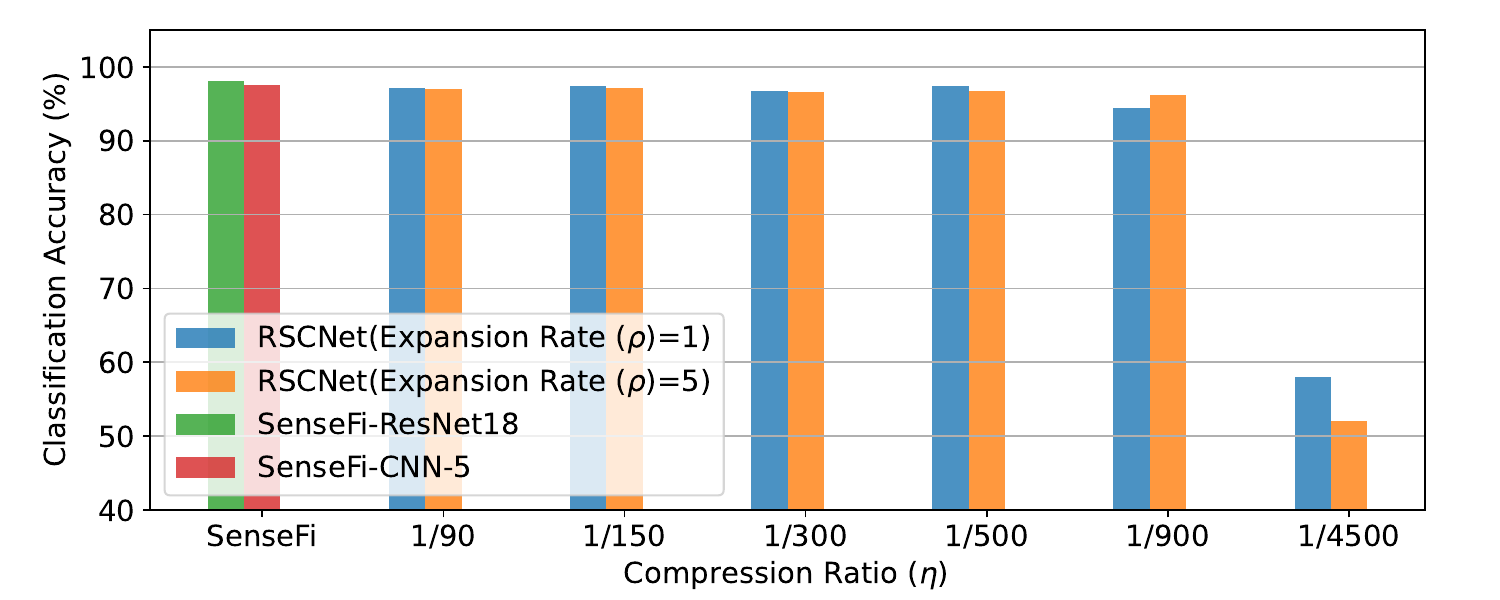}
        \caption{}
        \label{fig:accuracy}
    \end{subfigure}
    \begin{subfigure}{0.45\textwidth}
        \centering
        \includegraphics[width=\linewidth]{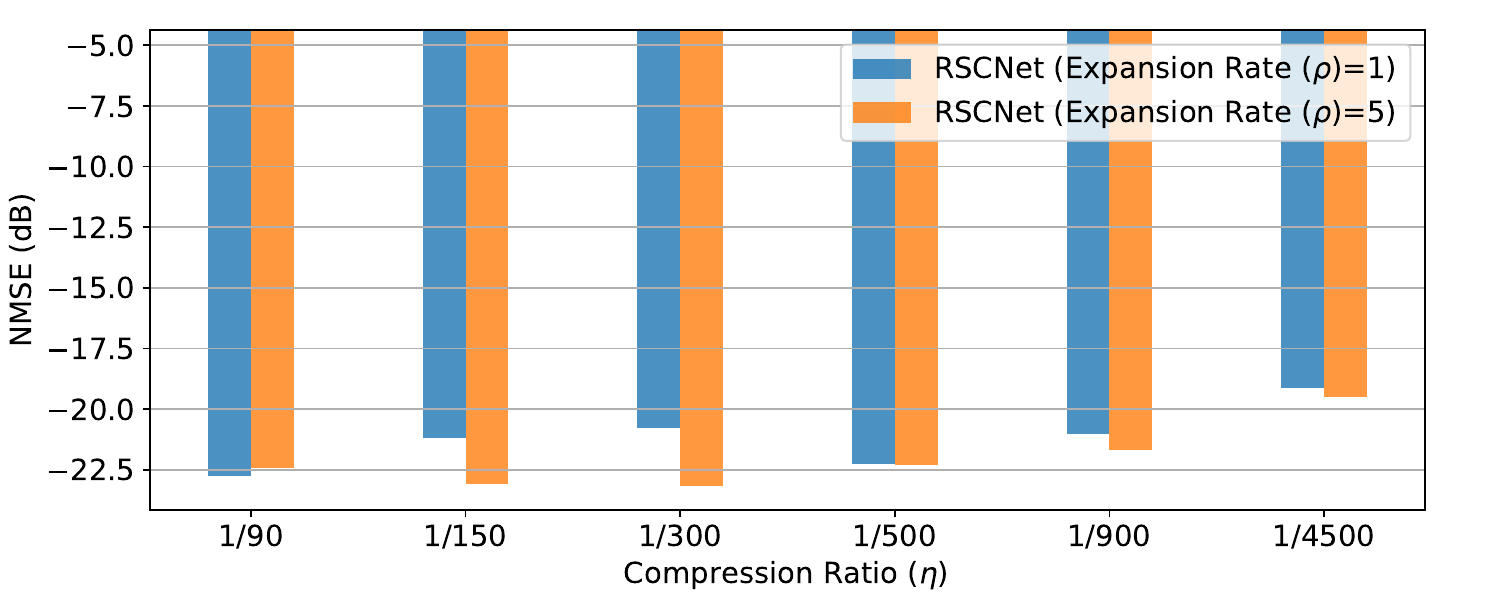}
        \caption{}
        \label{fig:nmse}
    \end{subfigure}
    \caption{Comparing performance metrics across RSCNet with different compression ratios and expansion rates for $N_f = 50$ as well as baseline methods, with (\subref{fig:accuracy}) illustrating sensing performance and (\subref{fig:nmse}) showcasing reconstruction error.}
    \label{fig:plot}
\end{figure}
\subsubsection{Compression Ratio vs NMSE and HAR Accuracy} The compression ratio, \( \eta \), is primarily contingent on the data overhead the communication channel with the cloud can accommodate. Nonetheless, it is crucial to optimize \( \eta \) to ensure a reasonable HAR  accuracy and CSI reconstruction performance (NMSE). Fig.~\ref{fig:plot} displays the RSCNet framework with different compression ratios for the decoder. Our primary objective, HAR, is benchmarked against the top two models from SenseFi~\cite{yang2023sensefi}. The HAR accuracy remains competitive with the SenseFi benchmarks, trailing by a mere 1-2\%, except at the most extreme compression ratio of \( \eta = 1/4500 \), which reduces a 50-frame CSI window to a single value. However, CSI reconstruction remains consistent across different \( \eta \) values, indicating the NMSE's reliability in various settings. 
\begin{figure*}[!htb]
    \centering
    \begin{subfigure}{0.24\linewidth}
        \centering
        \includegraphics[width=1\linewidth]{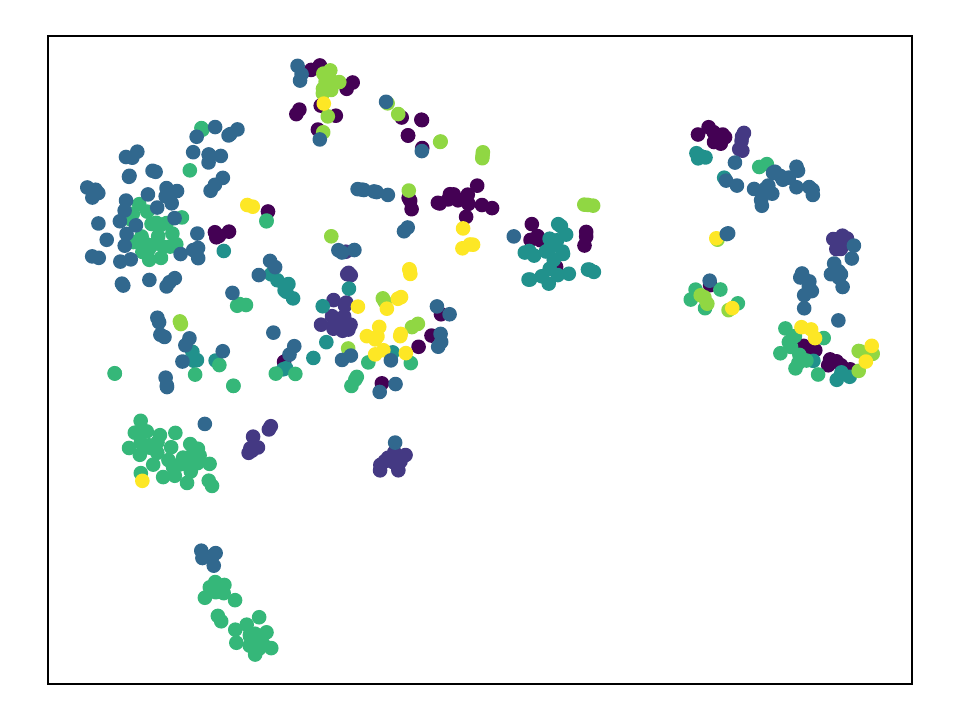}
        \caption{}
        \label{fig:raw}
    \end{subfigure}
    \begin{subfigure}{0.24\linewidth}
        \centering
        \includegraphics[width=1\linewidth]{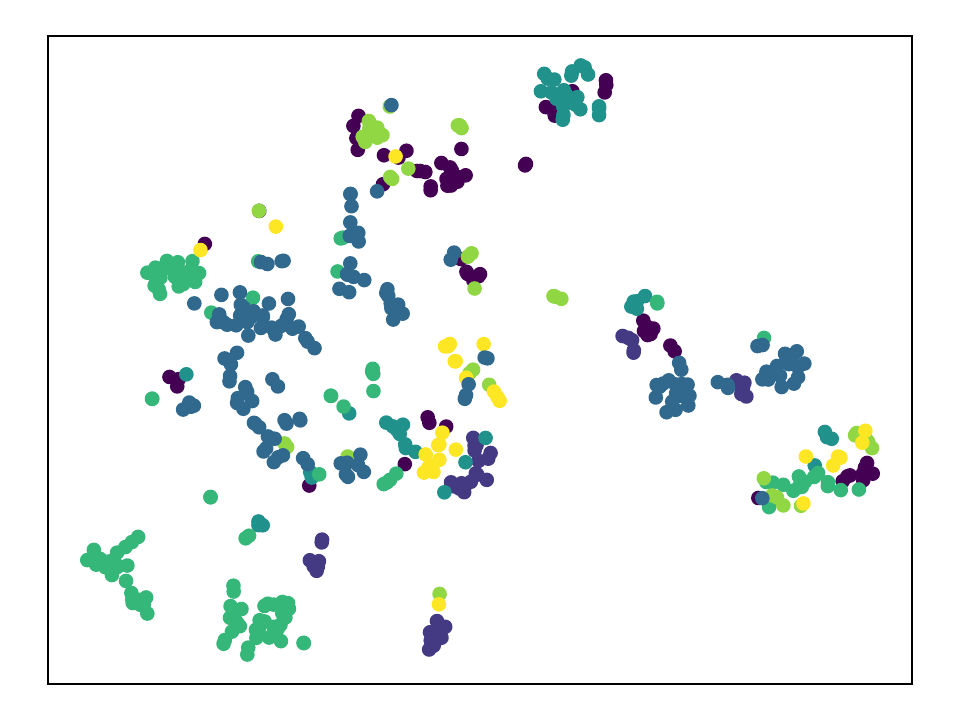}
        \caption{}
        \label{fig:compressed}
    \end{subfigure}
    \begin{subfigure}{0.24\linewidth}
        \centering
        \includegraphics[width=1\linewidth]{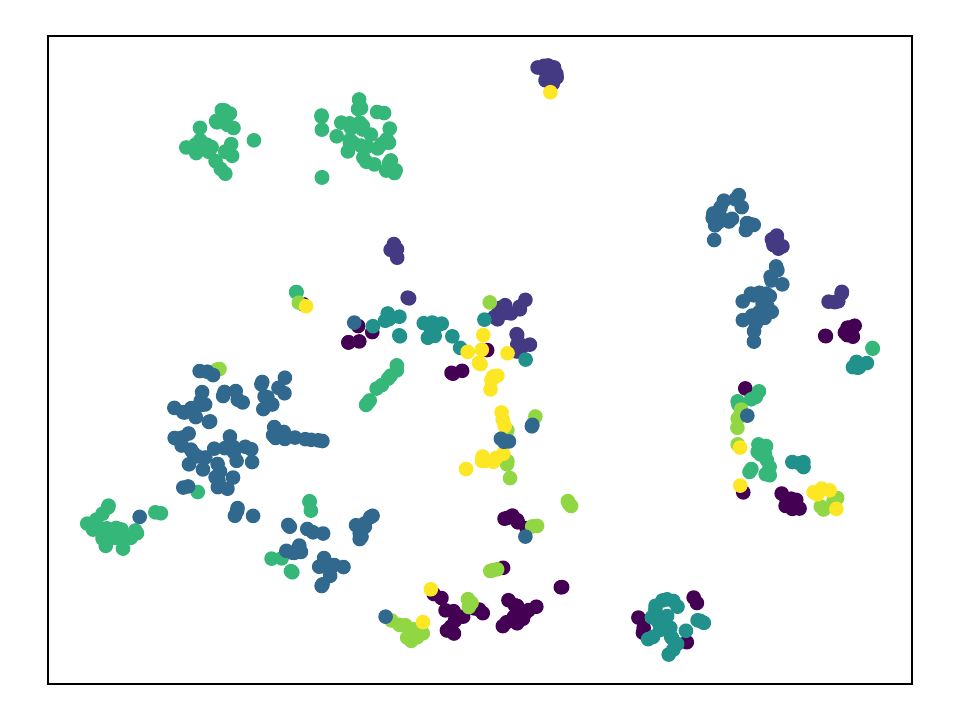}
        \caption{}
        \label{fig:recurrent}
    \end{subfigure}
    \begin{subfigure}{0.24\linewidth}
        \centering
        \includegraphics[width=1\linewidth]{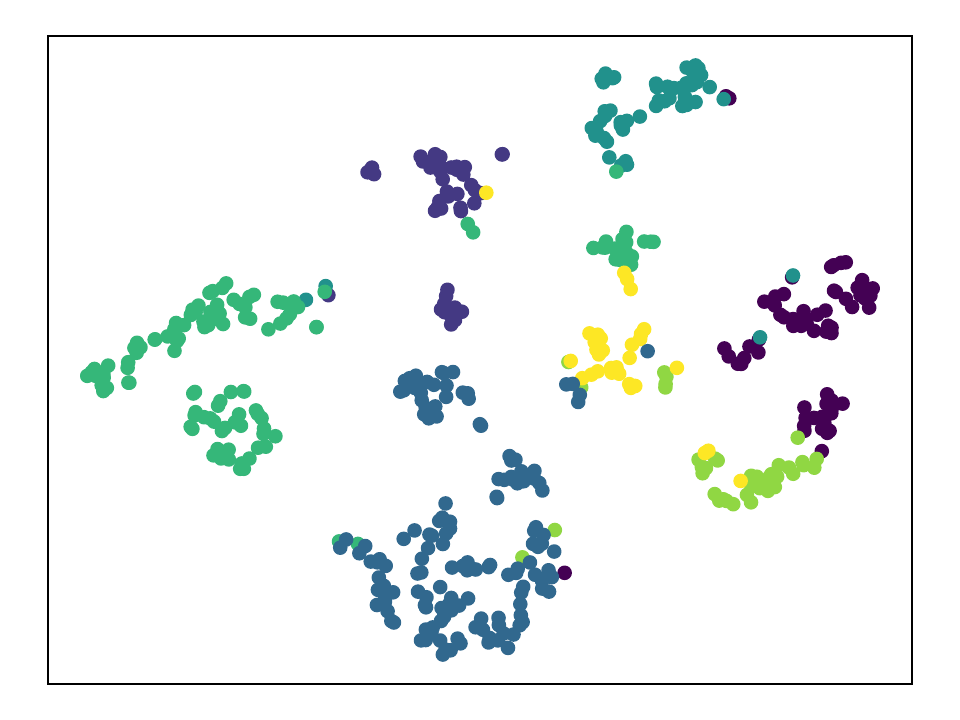}
        \caption{}
        \label{fig:classifier}
    \end{subfigure}
    \vspace{-2mm}
    \caption{Visualization via t-SNE for parameter settings $\eta=1/500$ and $N_f=50$ frames. (\subref{fig:raw}) Initial raw CSI representation; (\subref{fig:compressed}) Compressed CSI embedding; (\subref{fig:recurrent}) LSTM layer output embedding; (\subref{fig:classifier}) Final layer embedding within the classifier}
    \label{fig:tsne} 
\end{figure*}

\subsubsection{Expansion Rate vs NMSE}Fig.~\ref{fig:plot} displays the RSCNet framework with expansion rates \( \rho=1,5 \) for the decoder.
The RSCNet's reconstruction error is intrinsically linked to its decoder's expansion ratio, \( \rho \). As \( \rho \) rises, the decoder's convolutional layers broaden their channel dimensions, boosting the CSI data reconstruction ability. As depicted in Fig.~\ref{fig:flops}, the increased performance of the decoder is counteracted by heightened computational requirements. For instance, a decoder at \( \rho=5 \) can demand up to tenfold the computational resources than its \( \rho=1 \) counterpart, contingent on \( N_f \). Still, varying expansion rates offer a trade-off between reconstruction performance and computational efficiency, furnishing multiple deployment options for the decoder based on the application and the cloud service's resources. However, it is noteworthy that the encoder has a significantly lower computational cost than the decoder which makes it optimized for development on resource-limited devices.

\subsubsection{T-SNE Analysis} In Fig.~\ref{fig:tsne}, we present a t-SNE visualization of the CSI at various network phases, differentiated by the labels of corresponding activities. The visualization of raw CSI is illustrated in Fig.~\ref{fig:raw}. Although there is a discernible separation among different activity types, the majority of samples appear dispersed throughout the embedding space. Conversely, Fig.~\ref{fig:compressed} showcases the spatial separation of the compressed CSI, which is relayed to the cloud server. Notably, there is an enhanced distinction in the feature space. This indicates that the encoder serves a dual purpose: facilitating CSI reconstruction and being effectively discriminative for the HAR task. Despite the refined spatial delineation, overlaps across certain classes within the compressed CSI remain apparent. Advancing to the LSTM layer's representation in Fig. \ref{fig:recurrent}, a heightened discrimination towards class labels is evident, reflecting the LSTM's capability in capturing temporal features necessary for HAR. In our final representation, we delineate clusters of CSI in the classifier's concluding layer, revealing a distinct clustering contingent upon the activities. In summation, our visualizations highlight the encoder's role in both data compression and task-relevant feature discrimination, while also validating the LSTM block's effectiveness in temporal feature extraction. The ensuing representations manifest discernible, activity-centric clusterings, thereby substantiating our methodological propositions.

\section{Conclusion}
This paper proposed Real-time Sensing and Compression Network (RSCNet) which facilitates real-time compression and sensing through adaptable small CSI windows. RSCNet incorporates an LSTM block to enhance accuracy and reconstruction by using previous CSI windows information. Our evaluations consistently underline RSCNet's effectiveness, highlighting its compatibility with different CSI window sizes, and compression ratios, and comparing its performance  with state-of-the-art counterparts. 

\bibliography{ref}

\begin{thebibliography}{10}
\providecommand{\url}[1]{#1}
\csname url@samestyle\endcsname
\providecommand{\newblock}{\relax}
\providecommand{\bibinfo}[2]{#2}
\providecommand{\BIBentrySTDinterwordspacing}{\spaceskip=0pt\relax}
\providecommand{\BIBentryALTinterwordstretchfactor}{4}
\providecommand{\BIBentryALTinterwordspacing}{\spaceskip=\fontdimen2\font plus
\BIBentryALTinterwordstretchfactor\fontdimen3\font minus
  \fontdimen4\font\relax}
\providecommand{\BIBforeignlanguage}[2]{{%
\expandafter\ifx\csname l@#1\endcsname\relax
\typeout{** WARNING: IEEEtran.bst: No hyphenation pattern has been}%
\typeout{** loaded for the language `#1'. Using the pattern for}%
\typeout{** the default language instead.}%
\else
\language=\csname l@#1\endcsname
\fi
#2}}
\providecommand{\BIBdecl}{\relax}
\BIBdecl

\bibitem{6g}
M.~Rasti, S.~K. Taskou, H.~Tabassum, and E.~Hossain, ``Evolution toward 6g
  multi-band wireless networks: A resource management perspective,'' \emph{IEEE
  Wireless Communications}, vol.~29, no.~4, pp. 118--125, 2022.

\bibitem{9831898}
J.~Hu, J.~Yang, J.-B. Ong, D.~Wang, and L.~Xie, ``{ResFi}: {WiFi}-enabled
  device-free respiration detection based on deep learning,'' in \emph{2022
  IEEE 17th Intl. Conf. on Control \& Automation (ICCA)}, 2022, pp. 510--515.

\bibitem{8397121}
R.~Zhou, M.~Hao, X.~Lu, M.~Tang, and Y.~Fu, ``Device-free localization based on
  {CSI} fingerprints and deep neural networks,'' in \emph{2018 15th Annual IEEE
  Int. Conf. on Sensing, Commun., and Networking (SECON)}, 2018, pp. 1--9.

\bibitem{9834923}
Z.~Shi, Q.~Cheng, J.~A. Zhang, and R.~Yi~Da~Xu, ``Environment-robust
  {WiFi}-based human activity recognition using enhanced {CSI} and deep
  learning,'' \emph{IEEE Internet Things J.}, vol.~9, no.~24, pp.
  24\,643--24\,654, 2022.

\bibitem{9233449}
Y.~Zhao, R.~Gao, S.~Liu, L.~Xie, J.~Wu, H.~Tu, and B.~Chen, ``Device-free
  secure interaction with hand gestures in {WiFi}-enabled {IoT} environment,''
  \emph{IEEE Internet Things J.}, vol.~8, no.~7, pp. 5619--5631, 2021.

\bibitem{10.1145/1925861.1925870}
\BIBentryALTinterwordspacing
D.~Halperin, W.~Hu, A.~Sheth, and D.~Wetherall, ``Tool release: Gathering
  802.11n traces with channel state information,'' \emph{SIGCOMM Comput.
  Commun. Rev.}, vol.~41, no.~1, p.~53, jan 2011. [Online]. Available:
  \url{https://doi.org/10.1145/1925861.1925870}
\BIBentrySTDinterwordspacing

\bibitem{yang2023sensefi}
J.~Yang, X.~Chen, H.~Zou, C.~X. Lu, D.~Wang, S.~Sun, and L.~Xie, ``{SenseFi}: A
  library and benchmark on deep-learning-empowered {WiFi} human sensing,''
  \emph{Patterns}, vol.~4, no.~3, 2023.

\bibitem{9900419}
S.~M. Hernandez and E.~Bulut, ``{WiFi} sensing on the edge: Signal processing
  techniques and challenges for real-world systems,'' \emph{IEEE Commun. Surv.
  Tutor.}, vol.~25, no.~1, pp. 46--76, 2023.

\bibitem{zhuravchak2022human}
A.~Zhuravchak, O.~Kapshii, and E.~Pournaras, ``Human activity recognition based
  on {WiFi} {CSI} data - a deep neural network approach,'' \emph{Procedia
  Computer Science}, vol. 198, pp. 59--66, 2022.

\bibitem{9759238}
Y.~Gu, X.~Zhang, Y.~Wang, M.~Wang, H.~Yan, Y.~Ji, Z.~Liu, J.~Li, and M.~Dong,
  ``{WiGRUNT}: {WiFi}-enabled gesture recognition using dual-attention
  network,'' \emph{IEEE Transactions on Human-Machine Systems}, vol.~52, no.~4,
  pp. 736--746, 2022.

\bibitem{9516988}
Y.~Zhang, Y.~Zheng, K.~Qian, G.~Zhang, Y.~Liu, C.~Wu, and Z.~Yang,
  ``{Widar3.0}: Zero-effort cross-domain gesture recognition with {Wi-Fi},''
  \emph{IEEE TPAMI}, vol.~44, no.~11, pp. 8671--8688, 2022.

\bibitem{9797871}
S.~Tang, J.~Xia, L.~Fan, X.~Lei, W.~Xu, and A.~Nallanathan, ``Dilated
  convolution based {CSI} feedback compression for massive {MIMO} systems,''
  \emph{IEEE Trans. Veh. Tech.}, vol.~71, no.~10, pp. 11\,216--11\,221, 2022.

\bibitem{7207365}
M.~Hassanalieragh, A.~Page, T.~Soyata, G.~Sharma, M.~Aktas, G.~Mateos,
  B.~Kantarci, and S.~Andreescu, ``Health monitoring and management using
  internet-of-things ({IoT}) sensing with cloud-based processing: Opportunities
  and challenges,'' in \emph{2015 IEEE Int. Conf. on Services Computing}, 2015,
  pp. 285--292.

\bibitem{yang2022efficientfi}
J.~Yang, X.~Chen, H.~Zou, D.~Wang, Q.~Xu, and L.~Xie, ``Efficientfi: Toward
  large-scale lightweight {WiFi} sensing via {CSI} compression,'' \emph{IEEE
  Internet Things J.}, vol.~9, no.~15, pp. 13\,086--13\,095, 2022.

\bibitem{8067693}
S.~Yousefi, H.~Narui, S.~Dayal, S.~Ermon, and S.~Valaee, ``A survey on behavior
  recognition using {WiFi} channel state information,'' \emph{IEEE Commun.
  Mag.}, vol.~55, no.~10, pp. 98--104, 2017.

\bibitem{10.1145/3310194}
\BIBentryALTinterwordspacing
Y.~Ma, G.~Zhou, and S.~Wang, ``{WiFi} sensing with channel state information: A
  survey,'' \emph{ACM Comput. Surv.}, vol.~52, no.~3, June 2019. [Online].
  Available: \url{https://doi.org/10.1145/3310194}
\BIBentrySTDinterwordspacing

\end{thebibliography}
\bibliographystyle{IEEEtran}

\end{document}